\documentclass[preprintnumbers,showpacs,amsmath,amssymb,floatfix,prd,onecolumn,superscriptaddress,nofootinbib]{revtex4-2}

\usepackage{graphicx}
\usepackage{epsfig}
\usepackage{bm}
\usepackage{amsfonts}
\usepackage{epstopdf}
\usepackage[colorlinks,linkcolor=blue,anchorcolor=blue,citecolor=green,backref]{hyperref}
\usepackage{cancel}

\begin{document}
	
\title{Analytic Solution for the Motion of Spinning Particles in Plane Gravitational Wave Spacetime}

\author{Ke Wang}
\email[E-mail: ]{01wangke@mail.bnu.edu.cn}
\affiliation{School of Physics and Astronomy, Beijing Normal University, Beijing 100875, China}



\begin{abstract}
The interaction between spin and gravitational waves causes spinning bodies to deviate from their geodesics.  
In this work, we obtain the analytic solution of the 
Mathisson--Papapetrou--Dixon equations at linear order in the spin for plane gravitational wave spacetimes.
Our approach combines a parallel-transported tetrad with the translational 
Killing symmetries of plane wave spacetimes, yielding six conserved 
quantities that fully determine the momentum, spin evolution, and 
worldline.
The resulting transverse and longitudinal motions are expressed in closed 
form as single integrals of the retarded time, providing a unified and 
model-independent framework for computing spin--curvature-induced 
deviations.
This analytic solution offers a versatile tool for studying spin-dependent 
effects in gravitational memory, Penrose-limit geometries, and high-energy scattering regimes.
\end{abstract}


\maketitle
\tableofcontents

\section{Introduction}

The motion of spinning test particles in curved spacetime plays a central
role in many problems in general relativity.  
At the dipole level, the dynamics is governed by the
Mathisson--Papapetrou--Dixon (MPD) equations \cite{dixon_dynamics_1970,Papapetrou:1951pa}, which encode the coupling
between the particle's intrinsic spin and the background curvature.
This spin--curvature interaction generates corrections to geodesic motion,
influencing precession and orbital phasing.
Such effects are essential in a wide range of relativistic scenarios
\cite{Suzuki:1996gm,Dietrich:2020eud,semerak_spinning_1999,harte_extended-body_2020,Bini:2017xzy,Costa:2017kdr,Garriga:1990dp,Kessari:2002jc,andrzejewski_revisiting_2025}.

In many physically relevant situations, gravitational radiation can be
accurately modeled as a \emph{plane gravitational wave}.  
This description naturally emerges in the local wave zone of an isolated
source, where the curvature radius of the wavefront is much larger than
the scale of a detector or a test body, rendering the wavefront locally
planar. 
Moreover, plane wave spacetimes appear universally as \emph{Penrose limits} \cite{Penrose:1976} of
arbitrary geometries taken along null geodesics, providing a powerful local
description of the geometry experienced by highly boosted observers. 
These fundamental geometric features make plane gravitational waves an ideal theoretical
setting in which to study spin--curvature effects.

Solving the MPD equations has long been a significant challenge. The MPD equations possess sufficient generality, being applicable not only to the description of massive particles but also, in an approximate sense, to massless spinning particles arising in the high-frequency limit of field theories. In this paper, we focus primarily on the dynamics of massive spinning particles.
Recent work has shown that analytic MPD solutions can be obtained
at linear order in the spin in highly symmetric geometries such as
spherically symmetric black holes~\cite{witzany_analytic_2024,Ciou:2025ygb} and certain axisymmetric black holes~\cite{skoupy_analytic_2025,Chen:2025ncm}.
Previous works have also attempted to solve the MPD equations in plane wave spacetimes, yielding interesting results under different choices of spin supplementary conditions (SSC)~\cite{bini_gyroscopes_2000,bini_dixons_2009,mohseni_gravitational_2000,mohseni_motion_2001,bini_deviation_2017,Mohseni:2002zg}. It is worth noting that the spin Hall equations, a particular case of the MPD equations restricted to massless particles with longitudinal angular momentum, have recently been solved analytically in gravitational wave spacetimes~\cite{Harte:2024mwj}.

Spin--curvature dynamics in plane wave geometries is also closely tied to 
several modern developments in gravitational-wave physics.  
Recent studies have shown that freely falling gyroscopes may accumulate 
wave-induced precession~\cite{Seraj:2022qyt}, and that relative spin 
orientations can also encode memory-like signatures in plane wave backgrounds~\cite{Wang:2023eqj,Chen:2025tok}.  
These insights motivate the search for observable signatures of spin-gravitational-wave coupling in
future space-based detectors such as LISA, TianQin, or Taiji~\cite{LISA:2017pwj,LISA:2024hlh,TianQin:2015yph,Ruan:2018tsw}.
Owing to their exceptionally long interferometric baselines and highly stable orbital configurations,
such missions provide a natural and ideal environment for detecting spin-induced effects;
Consequently, a thorough understanding of spin--curvature dynamics in plane wave spacetimes is essential—not only for
advancing the theoretical framework of gravitational memory observables~\cite{Flanagan:2018yzh,Flanagan:2019ezo},
but also for assessing potential observational imprints in next-generation high-precision gravitational-wave experiments.

In this work, we focus on the leading-order spin--curvature coupling,
i.e., the linear-spin approximation $\mathcal{O}(s)$.  
This approximation is well justified in many physical contexts.
For realistic test bodies—such as spacecraft proof masses, gyroscopes~\cite{everitt_gravity_2011},
or small compact objects~\cite{Skoupy:2021asz}---the dimensionless ratio $s/m$ is typically
small, and higher-order spin multipoles are negligible.
At $\mathcal{O}(s)$, the MPD equations simplify considerably: the
momentum and velocity coincide, the spin magnitude is conserved, and the
curvature coupling reduces to a single linear term.  
The resulting system captures the dominant spin-dependent corrections
while retaining analytic tractability.

In this work, we provide a complete analytic solution of the MPD equations at
linear order in the spin, $\mathcal{O}(s)$, in the plane gravitational wave spacetime.
Our approach combines two key ingredients. First, we construct an orthonormal
tetrad that is explicitly parallel transported along geodesics.
Projecting the spin vector onto this tetrad yields three conserved
quantities at $\mathcal{O}(s)$, supplementing the three Killing constants
associated with the isometric symmetries of the plane wave spacetime.  
Taken together, these six constants fully determine the momentum and spin at
linear order.  
The resulting equations for the transverse momenta decouple naturally, allowing
the worldline components $y(u)$, $z(u)$ and the longitudinal coordinate $x(u)$
to be obtained by single integrals.

The paper is organized as follows.  
In Sec.~\ref{sec:MPD} we review the MPD equations under the Tulczyjew--Dixon SSC.  
Sec.~\ref{sec:spacetime} summarizes the geometry of plane waves and the associated Killing
constants.  
In Sec.~\ref{sec:tetrad} we construct the parallel-transported tetrad and identify the
spin-induced constants of motion.  
Sec.~\ref{sec:solution} presents the complete $\mathcal{O}(s)$ solution of the MPD system and the
resulting analytic worldline.  
We conclude in Sec.~\ref{sec:discussion} with a discussion of implications and possible future
directions.

\noindent\textbf{Notation.}
Throughout this work we employ natural units $G=c=1$ and the metric
signature $(-,+,+,+)$.
We adopt the convention $\varepsilon^{0123}=+1$ for the Levi--Civita
tensor, with $\epsilon^{\mu\nu\alpha\beta}=\varepsilon^{\mu\nu\alpha\beta}/\sqrt{-g}$.

\section{Mathisson--Papapetrou--Dixon equations under the Tulczyjew--Dixon SSC}
\label{sec:MPD}

The motion of a classical spinning test particle in curved spacetime is described by the Mathisson--Papapetrou--Dixon (MPD) equations \cite{dixon_dynamics_1970,Papapetrou:1951pa}
\begin{align}
    \frac{Dp^\mu}{d\tau} &= -\frac12\, R^\mu{}_{\nu\alpha\beta}\, u^\nu S^{\alpha\beta},\\
    \frac{DS^{\mu\nu}}{d\tau} &= p^\mu u^\nu - p^\nu u^\mu ,
    \label{eq:MPD}
\end{align}
where \(u^\mu = dx^\mu / d\tau\) is the tangent to the representative worldline parametrized by the affine parameter \(\tau\),
\(R^\mu{}_{\nu\alpha\beta}\) is the Riemann tensor, \(p^\mu\) is the four--momentum, and \(S^{\mu\nu}\) is the antisymmetric spin tensor.  

To uniquely specify the worldline of an extended body, a spin--supplementary condition (SSC) must be imposed.  
Throughout this work, we adopt the Tulczyjew--Dixon (TD) SSC \cite{Tulczyjew:1959,dixon_dynamics_1970}
\begin{equation}
    S^{\mu\nu} p_\nu = 0 ,
    \label{eq:TD-SSC}
\end{equation}
which guarantees that the momentum is timelike and that the mass
\(\mathcal{M}=\sqrt{-p_\mu p^\mu}\) is conserved.
The mass \(m=-p_\mu u^\mu\), however, is not conserved in general.

It is convenient to introduce the spin four-vector
\begin{equation}
    s^\mu \equiv -\frac{1}{2\mathcal{M}^2}\,
        \epsilon^{\mu\nu\alpha\beta} p_\nu S_{\alpha\beta},
    \label{eq:spin-vector}
\end{equation}
which satisfies \(s^\mu p_\mu = 0\) and whose magnitude
\(s^2=s_\mu s^\mu = S_{\mu\nu} S^{\mu\nu}/2\mathcal{M}^2\) is constant under TD SSC.
The inverse relation is \(S^{\mu\nu}=\epsilon^{\mu\nu\alpha\beta} p_\alpha s_\beta\).

Under the TD SSC the four--velocity \(u^\mu\) and the four--momentum \(p^\mu\) do not coincide.  
Although an exact momentum--velocity relation is known \cite{Tod:1976ud}, in this work we keep only the terms linear in the spin, \(\mathcal{O}(s)\).  
Standard manipulations yield
\begin{equation}
    p^\mu = m u^\mu + \mathcal{O}(s^2),
    \label{eq:U-P-relation}
\end{equation}
so that, at linear order in spin, the two different masses are equal, \(m=\mathcal{M}\).  
Moreover, the MPD equations \eqref{eq:MPD} reduce to
\begin{align}
    \frac{Du^\mu}{d\tau} &=
    -\frac{1}{2} R^\mu{}_{\nu\alpha\beta} u^\nu
      \epsilon^{\alpha\beta\kappa\lambda} u_\kappa s_\lambda
    + \mathcal{O}(s^2),\\
    \frac{DS^{\mu\nu}}{d\tau} &= 0 + \mathcal{O}(s^2).
    \label{eq:MPD-linear}
\end{align}
Using the relation \(S^{\mu\nu}=\epsilon^{\mu\nu\alpha\beta} p_\alpha s_\beta\), one further obtains
\[
    \frac{Ds^\mu}{d\tau}
    = \dot{u}^\nu u^\mu s_\nu,
\]
which corresponds precisely to the Fermi--Walker transport of the spin vector~\cite{poisson_motion_2011} (the missing item \(\dot{u}^\mu u^\nu s_\nu\) vanishes due to TD SSC).

When the background spacetime admits Killing symmetries, the corresponding Killing vectors generate conserved quantities.
For any Killing vector \( \xi^\mu \), the MPD dynamics preserves the quantity \cite{rudiger_conserved_1981,rudiger_conserved_1983,dixon_dynamics_1970}
\begin{equation}
    \mathcal{J}_\xi
    = \xi_\mu p^\mu
      - \frac12\, S^{\mu\nu} \nabla_\nu \xi_\mu ,
    \label{eq:Killing-constant}
\end{equation}
which is conserved along the MPD trajectory to all orders in spin.  
These integrals of motion reduce the effective dimensionality of the system.
As we shall see, the plane gravitational wave spacetime geometry introduced in the next section admits three translational Killing vectors that directly yield conserved components of the momentum, allowing all \(\mathcal{O}(s)\) corrections to be expressed in closed form.

In the following sections, we apply this formalism to the plane gravitational wave spacetime, construct a parallel-transported tetrad along a fiducial reference geodesic, and use it to obtain an analytic \(\mathcal{O}(s)\) solution of the MPD equations.

\section{Plane gravitational wave spacetime and geodesics}
\label{sec:spacetime}

We consider a plane gravitational wave propagating in the
$x$--direction, described in the standard null coordinates
\((u,v,y,z)\) by the metric
\begin{equation}
    ds^2 = 2\, du\, dv
    + (1 - h_{+})\, dy^2
    + (1 + h_{+})\, dz^2
    - 2 h_{\times}\, dy\, dz ,
    \label{eq:metric-plane-wave}
\end{equation}
where \(u=(t-x)/\sqrt{2}\) is the retarded time and
\(v=-(t+x)/\sqrt{2}\) is the advanced time.
The functions \(h_{+}(u)\) and \(h_{\times}(u)\) represent,
respectively, the $+$ and $\times$ polarization modes of the gravitational wave.
The determinant of the metric is denoted by 
\(\Delta \equiv |g|\), explicitly given by
\begin{equation}
    \Delta = 1 - h_{+}^2(u) - h_{\times}^2(u).
\end{equation}
It is imperative to clarify the geometric and physical nature of this metric. Provided that the functions $h_{+}(u)$ and $h_{\times}(u)$ are strictly constrained to satisfy the exact vacuum Einstein field equations ($R_{\mu\nu}=0$), this metric constitutes a specific exact plane wave spacetime in standard Rosen coordinates \cite{Einstein:1937qu}. Under this stringent nonlinear geometric constraint, the functions cannot represent completely arbitrary wave profiles, and the solutions are valid within the local coordinate chart prior to the formation of coordinate singularities.

However, a significant advantage of this formulation is its seamless connection to the weak-field regime. When analyzing realistic gravitational waves under the linear approximation in the transverse-traceless (TT) gauge ($|h| \ll 1$), the exact vacuum constraints relax. In this limit, $h_{+}(u)$ and $h_{\times}(u)$ recover their physical interpretations as arbitrary $+$ and $\times$ polarization modes. Therefore, the exact analytical solutions derived in this framework can be directly retained and applied up to the leading order of the wave amplitude, $\mathcal{O}(h)$.

\subsection{Killing symmetries and conserved momenta}

The plane wave metric \eqref{eq:metric-plane-wave} admits three
translational Killing vectors,
\begin{equation}
    \xi_{(v)}^\mu = \partial_v ,\qquad
    \xi_{(y)}^\mu = \partial_y ,\qquad
    \xi_{(z)}^\mu = \partial_z .
    \label{eq:Killing-vectors}
\end{equation}
Substituting these vectors into the general conserved quantity
\eqref{eq:Killing-constant}, one obtains three integrals of motion
for the MPD system.
The conserved quantity associated with \(\xi_{(v)}^\mu\) yields
\begin{equation}
    \mathcal{E} = p^u,
    \label{eq:constant-E}
\end{equation}
which plays the role of the ``longitudinal energy" in the $u$--direction.

The constants associated with \(\xi_{(y)}^\mu\) and \(\xi_{(z)}^\mu\) give the following expressions:
\begin{align}
    J_\alpha &= (1 - h_{+})\, p^y
           - h_{\times}\, p^z
           + \frac{p^{[z}s^{u]}}{\sqrt{\Delta}}
             \Big[
                h_{\times}\, \dot{h}_{\times}
                + (1 + h_{+})\, \dot{h}_{+}
             \Big] \nonumber \\
        &\quad
          - \frac{p^{[y}s^{u]}}{\sqrt{\Delta}}
             \Big[
                h_{\times}\, \dot{h}_{+}
                + (1 - h_{+})\, \dot{h}_{\times}
             \Big] , 
    \label{eq:Jy}
\\[6pt]
    J_\beta &= (1 + h_{+})\, p^z
           - h_{\times}\, p^y
           + \frac{p^{[z}s^{u]}}{\sqrt{\Delta}}
             \left[
                (1 + h_{+})\, \dot{h}_{\times}
                - h_{\times}\, \dot{h}_{+}
             \right] \nonumber \\
        &\quad
          + \frac{p^{[y}s^{u]}}{\sqrt{\Delta}}
             \left[
                (1 - h_{+})\, \dot{h}_{+}
                - h_{\times}\, \dot{h}_{\times}
             \right] ,
    \label{eq:Jz}
\end{align}
where dots denote derivatives with respect to the retarded time \(u\),
and the antisymmetrization notation is defined as \(p^{[z}s^{u]} = (p^z s^u - p^u s^z)/2 \).
the two constants \( J_\alpha\) and \( J_\beta\) serve as the ``conserved momentum" in the $y$- and $z$-direction.

Equations \eqref{eq:Jy}--\eqref{eq:Jz} completely determine
the transverse components of the momentum once the spin vector is known.
In the absence of spin, i.e., in the strict limit \(\mathcal{O}(s^0)\),
the above integrals immediately imply
\begin{equation}
    p^y_{(0)} = \frac{(1+h_+) J_\alpha^{(0)} + h_{\times} J_\beta^{(0)}}{\Delta},\qquad
    p^z_{(0)} = \frac{h_{\times} J_\alpha^{(0)} + (1-h_+) J_\beta^{(0)}}{\Delta},
    \label{eq:p-geodesic}
\end{equation}
where the index (0) denotes the order of \(\mathcal{O}(s^0)\). Substituting these expressions into the normalization condition
\(u^\mu u_\mu = -1\) yields the geodesic equations \cite{Wang:2024dmn}.
The resulting worldline \(x^\mu_{(0)}(\tau)\) coincides with the geodesic solution in~\cite{Wang:2024dmn}.

In the next section, we proceed beyond the geodesic order and incorporate the \(\mathcal{O}(s)\) spin--curvature coupling.
A parallel-transported tetrad will be constructed along a fiducial
geodesic, allowing the full MPD system to be solved analytically.

\section{Parallel-transported tetrad in plane gravitational wave spacetime}
\label{sec:tetrad}

As discussed in Sec.~\ref{sec:spacetime}, the conserved quantities 
\((\mathcal{E},J_\alpha,J_\beta)\) obtained from the Killing symmetries of the
plane wave metric are not sufficient to algebraically determine all
components of the momentum in the MPD system.
In contrast to geodesic motion, the spin--curvature force introduces
additional degrees of freedom through the spin vector \(s^\mu\), and
one requires extra conserved quantities of order \(\mathcal{O}(s)\)
to achieve integrability.

Following the strategy developed by \citet{skoupy_analytic_2025} and \citet{marck_parallel-tetrad_1983},
we construct a parallel--transported orthonormal tetrad along a fiducial
reference geodesic.  
Expressing the spin vector on this tetrad basis yields three 
\(\mathcal{O}(s)\) constants of motion, which will supply the remaining
information necessary to solve the MPD equations in closed form.

\subsection{Construction of an orthonormal tetrad along a worldline}

Let \(u^\mu=(u^0,u^1,u^2,u^3)\) denote the tangent vector of a general timelike worldline, \(u^\mu u_\mu=-1\).
We take the zeroth leg of the tetrad to be
\begin{equation}
    e_0^\mu = u^\mu ,
\end{equation}
normalized by construction.
A remarkable property of the plane gravitational wave spacetime
is the existence of a covariantly constant null vector field \(l^\mu = \partial_v\) \cite{Blau:2002js},
satisfying \(\nabla_\alpha l^\mu = 0\).  
This allows us to construct the first spatial leg as
\begin{equation}
    e_1^\mu = \frac{1}{u^0}\,(\partial_v)^\mu + u^\mu,
    \label{eq:e1-def}
\end{equation}
which is orthogonal to \(u^\mu\) and has unit norm along the worldline.

To complete the spacelike basis in the transverse \((y,z)\)-plane,  
we introduce two preliminary orthonormal vectors
\(\tilde e_2^\mu\) and \(\tilde e_3^\mu\) that:
\begin{align}
    \tilde e_2^\mu &=
        \frac{1}{\sqrt{1-h_{+}}}\,(\partial_y)^\mu
        - \frac{u^3h_\times-u^2(1-h_+)}{u^0\,\sqrt{1-h_{+}}}\,
          (\partial_v)^\mu , \\[4pt]
    \tilde e_3^\mu &=
       \sqrt {\frac{1-h_{+}}{\Delta}}\,
        \big[(\partial_z)^\mu
        + \frac{h_{\times}}{1-h_{+}}\, (\partial_y)^\mu
        - \frac{u^3 \Delta}{u^0 (1-h_{+})}\,
          (\partial_v)^\mu \big],
\end{align}
where \(\Delta=1-h_+^2-h_\times^2\) as defined previously.
It is straightforward to verify that \(e_0^\mu,\quad e_1^\mu,\quad \tilde e_2^\mu,\quad \tilde e_3^\mu \)
form an orthonormal tetrad along the worldline \(u^\mu\).

Now let the worldline \(u^\mu\) be a geodesic \(u^\mu\to u^\mu_g\), then the legs \(e_0^\mu\) and \(e_1^\mu\) become parallel transported, 
but the vectors \(\tilde e_2^\mu\) and \(\tilde e_3^\mu\) are not parallel transported:
\begin{equation}
    u^\nu_g \nabla_\nu \tilde e_2^\mu \neq 0,
    \qquad
    u^\nu_g \nabla_\nu \tilde e_3^\mu \neq 0 .
\end{equation}

Because the transverse plane is two--dimensional, a single time-dependent
rotation is sufficient to convert the pair
\((\tilde e_2^\mu,\tilde e_3^\mu)\) into a parallel--transported pair.
We therefore define
\begin{align}
    e_2^\mu &= 
        \tilde e_2^\mu \cos\psi(u)
        - \tilde e_3^\mu \sin\psi(u), \label{eq:e2-def} \\[3pt]
    e_3^\mu &= 
        \tilde e_2^\mu \sin\psi(u)
        + \tilde e_3^\mu \cos\psi(u), \label{eq:e3-def}
\end{align}
and require that the parallel--transport conditions
\begin{equation}
    u^\nu_g \nabla_\nu e_2^\mu = 0,
    \qquad
    u^\nu_g \nabla_\nu e_3^\mu = 0
\end{equation}
hold along the reference geodesic \(u^\mu_g\).

A direct computation by expanding \( \tilde{e}_{\mu }^3u^{\nu}_g \nabla _{\nu }e_2^{\mu }=0\) yields the simple evolution equation
\begin{equation}
    \frac{d\psi}{du}
    = -\frac{h_{\times}\,\dot{h}_{+}
      + (1-h_{+})\,\dot{h}_{\times}}{2\,\sqrt{\Delta}(1-h_{+})} \, ,
    \label{eq:psi-evolution}
\end{equation}
which completely determines the rotation angle once the initial value \(\psi(0)\) is specified.
The final tetrad
\(
    \{e_0^\mu, e_1^\mu, e_2^\mu, e_3^\mu\}
\)
is thus orthonormal and parallel transported along the geodesic.

\subsection{Spin decomposition and conserved \texorpdfstring{$\mathcal{O}(s)$}{O(s)} quantities}

With the parallel--transported tetrad in hand, we can decompose the spin
four--vector as \( s^\mu= s_{\parallel}\, e_1^\mu+ s_{\alpha}\, e_2^\mu+ s_{\beta}\, e_3^\mu\), where
\begin{equation}
    s_{\parallel} = s_\mu e_1^\mu,\qquad
    s_{\alpha} = s_\mu e_2^\mu,\qquad
    s_{\beta} = s_\mu e_3^\mu .
    \label{eq:spin-conserved}
\end{equation}
Because each tetrad leg is parallel transported,
the coefficients
\((s_{\parallel},s_{\alpha},s_{\beta})\)
are constants of motion of order \(\mathcal{O}(s)\):
\begin{equation}
    \frac{ds_{\parallel}}{d\tau}
    = \frac{ds_{\alpha}}{d\tau}
    = \frac{ds_{\beta}}{d\tau}
    = 0 + \mathcal{O}(s^2).
\end{equation}

Together with the Killing constants 
\((\mathcal{E},J_\alpha,J_\beta)\),
these conserved spin components provide a complete set of algebraic
relations that allow all components of the momentum \(p^\mu\) to be
determined in closed form at order \(\mathcal{O}(s)\).
In the next section, we solve the MPD system explicitly.

\section{Solution at order \texorpdfstring{$\mathcal{O}(s)$}{O(s)}}
\label{sec:solution}

In this section, we combine the Killing constants
\((\mathcal{E},J_\alpha,J_\beta)\) introduced in Sec.~\ref{sec:spacetime}
with the spin constants
\((s_{\parallel},s_{\alpha},s_{\beta})\) defined in
Sec.~\ref{sec:tetrad} to obtain the explicit
\(\mathcal{O}(s)\) solution of the MPD equations.
We first derive the momentum and spin components in the
\((u,v,y,z)\) coordinate basis, and then rewrite the worldline as a set of
first--order equations in the retarded time \(u\).

\subsection{Momentum and spin components at \texorpdfstring{$\mathcal{O}(s)$}{O(s)}}

Inverting the decomposition
\(s^\mu = s_{\parallel} e_1^\mu
      + s_{\alpha} e_2^\mu
      + s_{\beta} e_3^\mu\),
one obtains the coordinate components of the spin vector as functions of \(u\):
\begin{align}
    s^u &= \mathcal{E}\, s_{\parallel}, 
    \label{eq:su-sol}\\[3pt]
    s^y &= 
      \frac{(1+h_+) J_\alpha + h_{\times} J_\beta}{\Delta}\, s_{\parallel}
      + \frac{s_{\alpha}\cos\psi + s_{\beta}\sin\psi}{\sqrt{1-h_+}}
      + \frac{h_{\times}}{\sqrt{\Delta(1-h_+)}}
        \big(\cos\psi\, s_{\beta} - \sin\psi\, s_{\alpha}\big),
    \label{eq:sy-sol}\\[3pt]
    s^z &=
      \frac{h_{\times} J_\alpha + (1-h_+) J_\beta}{\Delta}\, s_{\parallel}
      + \sqrt{\frac{1-h_+}{\Delta}}
        \big(-\sin\psi\, s_{\alpha} + \cos\psi\, s_{\beta}\big),
    \label{eq:sz-sol}
\end{align}
where the rotation angle \(\psi(u)\) is determined by
Eq.~\eqref{eq:psi-evolution} together with an initial value \(\psi(0)\).

Using the Killing integrals \eqref{eq:constant-E},
\eqref{eq:Jy} and \eqref{eq:Jz}, together with
Eqs.~\eqref{eq:su-sol}--\eqref{eq:sz-sol}, the \(\mathcal{O}(s)\)
corrections to the momentum components are obtained by solving a linear
system in \((p^y,p^z)\).
It is convenient to write the result in the compact form
\begin{align}
    p^u &= \mathcal{E}, 
    \label{eq:pu-sol}\\[3pt]
    p^y &= \frac{(1+h_+) J_\alpha + h_{\times} J_\beta}{\Delta} 
      + F^{(y)}_{u}(u)\, s^u
      + F^{(y)}_{y}(u)\, s^y
      + F^{(y)}_{z}(u)\, s^z,
      \label{eq:py-sol}\\[3pt]
    p^z &= \frac{h_{\times} J_\alpha + (1-h_+) J_\beta}{\Delta}
      + F^{(z)}_{u}(u)\, s^u
      + F^{(z)}_{y}(u)\, s^y
      + F^{(z)}_{z}(u)\, s^z,
      \label{eq:pz-sol}
\end{align}
where the coefficient functions 
\(F^{(A)}_{B}(u)\) (\(A=y,z\), \(B=u,y,z\)) are explicit functions of
\(h_+(u)\), \(h_\times(u)\), and their derivatives:
\begin{align}
    F^{(y)}_{u}(u) &=
    \frac{(1+h_+)\big(\dot{h}_{\times }J_{\alpha }+\dot{h}_+ J_{\beta }\big)
          -h_{\times }\big(\dot{h}_+ J_{\alpha }+\dot{h}_{\times }J_{\beta }\big)}
    {2 \Delta ^{3/2}},  \\[3pt]
    F^{(y)}_{y}(u) &=
    -\mathcal{E}\,
    \frac{\big(1+h_{\times }^2-h_+^2\big)\dot{h}_{\times }+2h_+ h_\times \dot{h}_+}
    {2 \Delta ^{3/2}},  \\[3pt]
    F^{(y)}_{z}(u) &=
    \mathcal{E}\,
    \frac{2h_\times (1+h_+)\dot{h}_{\times }
          -h_{\times }^2\dot{h}_+  + (1+h_+)^2 \dot{h}_+}
    {2 \Delta ^{3/2}},  \\[3pt]
    F^{(z)}_{u}(u) &=
    \frac{h_\times \big(\dot{h}_{\times }J_{\alpha }-\dot{h}_+ J_{\beta }\big)
          -(1-h_+)\big(\dot{h}_{\times }J_{\beta }+\dot{h}_+ J_{\alpha }\big)}
    {2 \Delta ^{3/2}},  \\[3pt]
    F^{(z)}_{y}(u) &=
    -\mathcal{E}\,
    \frac{2h_\times (1-h_+)\dot{h}_{\times }
          +h_{\times }^2\dot{h}_+-\dot{h}_+ (1-h_+)^2}
    {2 \Delta ^{3/2}},  \\[3pt]
    F^{(z)}_{z}(u) &=
    \mathcal{E}\,
    \frac{\big(1+h_{\times }^2-h_+^2\big)\dot{h}_{\times }
          +2 h_\times h_+ \dot{h}_+}
    {2 \Delta ^{3/2}}.
\end{align}

Equations \eqref{eq:su-sol}--\eqref{eq:sz-sol} and
\eqref{eq:pu-sol}--\eqref{eq:pz-sol} provide the explicit analytic
solution for the momentum and spin of the spinning particle at
order \(\mathcal{O}(s)\) in the plane gravitational wave spacetime.

\subsection{Worldline equations in retarded time}

In the Tulczyjew--Dixon SSC, at linear order in the spin we may identify
\(p^\mu = m u^\mu + \mathcal{O}(s^2)\), see Eq.~\eqref{eq:U-P-relation},
and the two mass definitions coincide, \(m=\mathcal{M}\).
From Eq.~\eqref{eq:pu-sol} we have
\begin{equation}
    \frac{du}{d\tau} = u^u = \frac{p^u}{m}
    = \frac{\mathcal{E}}{m},
\end{equation}
so that the retarded time \(u\) is an affine parameter along the
worldline.  
Using \(d/d\tau = (du/d\tau)\, d/du\), the transverse components of the
four--velocity can be written as
\begin{equation}
    \frac{dy}{du} = \frac{u^y}{u^u}
    = \frac{p^y}{p^u}
    = \frac{1}{\mathcal{E}}\, p^y(u),
    \qquad
    \frac{dz}{du} = \frac{u^z}{u^u}
    = \frac{p^z}{p^u}
    = \frac{1}{\mathcal{E}}\, p^z(u).
    \label{eq:dydu-dzdu}
\end{equation}
Since \(p^y(u)\) and \(p^z(u)\) depend only on the retarded time through
\(h_+(u)\), \(h_\times(u)\) and the constants
\((\mathcal{E},J_\alpha,J_\beta,s_{\parallel},s_{\alpha},s_{\beta})\),
the equations \eqref{eq:dydu-dzdu} are decoupled first--order ordinary
differential equations for \(y(u)\) and \(z(u)\).

Integrating once, we obtain the transverse worldline in closed form,
\begin{align}
    y(u) &= y(0)
      + \frac{1}{\mathcal{E}}
        \int_{0}^{u} p^y(u')\, du', \label{eq:y-solution}\\[3pt]
    z(u) &= z(0)
      + \frac{1}{\mathcal{E}}
        \int_{0}^{u} p^z(u')\, du'. \label{eq:z-solution}
\end{align}
In the geodesic limit, the integrands reduce to the spinless momenta
\(p^y_{(0)}(u)\) and \(p^z_{(0)}(u)\) given in Eq.~\eqref{eq:p-geodesic},
and Eqs.~\eqref{eq:y-solution}--\eqref{eq:z-solution} reproduce the
geodesics discussed in Sec.~\ref{sec:spacetime}.

The remaining longitudinal motion can be reconstructed from the
normalization condition \(u^\mu u_\mu=-1\) together with the
relations \(u=(t-x)/\sqrt{2}\) and \(v=-(t+x)/\sqrt{2}\).
Using the metric \eqref{eq:metric-plane-wave} and dividing
\(u_\mu u^\mu=-1\) by \((u^u)^2\) we find
\begin{equation}
    2\,\frac{dv}{du}
    + (1-h_+)\Big(\frac{dy}{du}\Big)^2
    + (1+h_+)\Big(\frac{dz}{du}\Big)^2
    - 2 h_{\times}\,\frac{dy}{du}\frac{dz}{du}
    = -\Big(\frac{m}{\mathcal{E}}\Big)^2 .
\end{equation}
In terms of the Cartesian coordinate \(x\), \(dx/du= -(1+dv/du)/\sqrt{2}\). By integrating once, we obtain
\begin{equation}
    x(u) = x(0)
      - \frac{u}{\sqrt{2}}
      + \frac{1}{2\sqrt{2}}\int_0^{u} du'\,
      \Bigg[
        (1-h_+(u'))\Big(\frac{dy}{du'}\Big)^2
        + (1+h_+(u'))\Big(\frac{dz}{du'}\Big)^2
        - 2 h_{\times}(u')\,\frac{dy}{du'}\frac{dz}{du'}
        + \Big(\frac{m}{\mathcal{E}}\Big)^2
      \Bigg].
    \label{eq:x-solution}
\end{equation}

Equations \eqref{eq:y-solution}, \eqref{eq:z-solution} and
\eqref{eq:x-solution} provide the complete \(\mathcal{O}(s)\) worldline of the spinning test particle in the plane
gravitational wave spacetime.

To obtain a concrete worldline, the constants 
\((\mathcal{E},J_\alpha,J_\beta,s_{\parallel},s_\alpha,s_\beta)\)
must be fixed by specifying initial conditions at \(u=0\).
Given the initial position \(x^\mu(0)\), the initial momentum \(p^\mu(0)\),
and the initial spin vector \(s^\mu(0)\), one may project the momentum
and spin onto the parallel-transported tetrad constructed in
Sec.~\ref{sec:tetrad}.  In particular,
\begin{itemize}
    \item \(\mathcal{E}=p^u(0)\) follows directly from Eq.~\eqref{eq:constant-E};
    \item \(J_\alpha\) and \(J_\beta\) follow from Eqs.~\eqref{eq:Jy}--\eqref{eq:Jz} evaluated at \(u=0\);
    \item \(s_\parallel,s_\alpha,s_\beta\) are obtained by projecting \(s^\mu(0)\) onto \(\{e_1,e_2,e_3\}\) at \(u=0\).
\end{itemize}
The only gauge freedom in this construction is the initial tetrad
rotation angle \(\psi(0)\), corresponding to a rigid rotation in the
transverse \((y,z)\)-plane.  For concreteness, we set \(\psi(0)=0\) in
the examples below.

\subsection{Examples: a static spinning particle}

To illustrate the spin-induced deviation from geodesic motion,
we consider a monochromatic plane wave,
\begin{equation}
    h_{+}(u)=h \sin u,\qquad 
    h_{\times}(u)=h \cos u,
\end{equation}
and a particle initially at rest at the origin, here \(h \) represents the amplitude of the gravitational wave.
For clarity, we decompose the spin into a longitudinal component,
\(s_\parallel\), aligned with the propagation direction,
and a transverse magnitude 
\(s_\perp=\sqrt{s_\alpha^2+s_\beta^2}\).

\paragraph*{Case I: purely longitudinal spin.}
If the initial spin is aligned with the propagation direction,
\(s_\perp=0\), then Eqs.~\eqref{eq:Jy}--\eqref{eq:Jz} give
\(J_\alpha=J_\beta=0\).
In addition, the terms proportional to \(s^u\) in 
Eqs.~\eqref{eq:py-sol}--\eqref{eq:pz-sol} vanish.
Thus all \(\mathcal{O}(s)\) corrections to the transverse momentum vanish
identically, and the resulting worldline coincides with the geodesic.
This confirms that a purely longitudinal spin does not couple to the
plane wave at linear order.

\begin{figure}[htbp]
    \centering
    \includegraphics[width=0.52\textwidth]{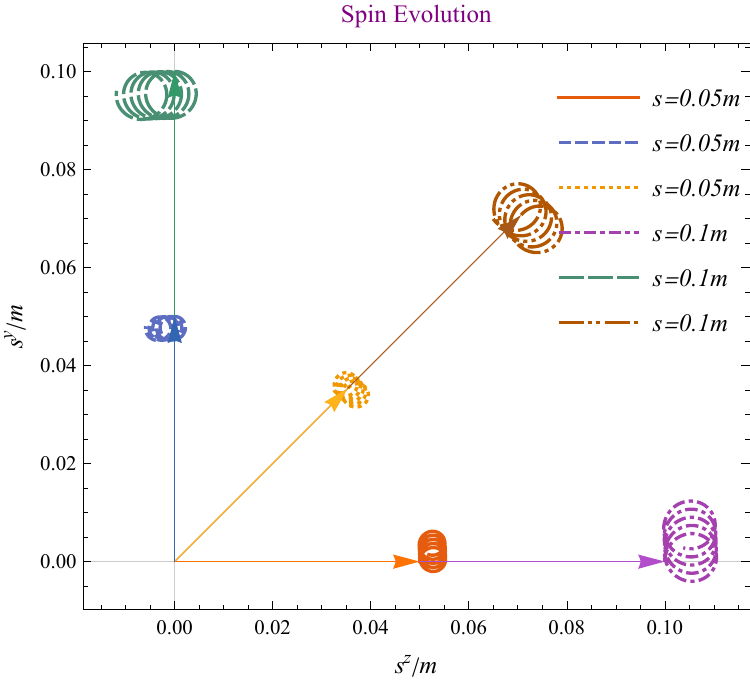}
    \caption{
    Evolution of the coordinate components \(s^y(u)\) and \(s^z(u)\) for a
    particle initially at rest, with transverse spin magnitude
    \(s_\perp = 0.1\,m\) and \(s_\perp = 0.05\,m\). The amplitude of the gravitational wave is \(h=10^{-1}\).
    Although the tetrad components of the spin remain constant, the 
    coordinate components undergo a precession driven by the plane wave
    geometry.  
    The arrows indicate the initial spin direction.}
    \label{fig:spinevolution}
\end{figure}

\paragraph*{Case II: transverse spin.}
When the particle carries transverse spin \(s_\perp\neq 0\), the
spin--curvature force becomes active and induces oscillatory corrections
to the momentum and trajectory.  
Figure~\ref{fig:spinevolution} shows the evolution of the coordinate
spin components \(s^y(u)\) and \(s^z(u)\) for particles with two 
different magnitudes, \(s_\perp=0.05\,m\) and \(s_\perp=0.1\,m\).
Although the tetrad components \((s_\parallel,s_\alpha,s_\beta)\) are
constant, the coordinate components exhibit a characteristic precession
pattern determined by the rotation angle \(\psi(u)\).
The larger the transverse spin, the larger the amplitude of the
precession ellipse traced out in the \((s^y,s^z)\) plane.

\begin{figure}[htbp]
    \centering
    \includegraphics[width=0.52\textwidth]{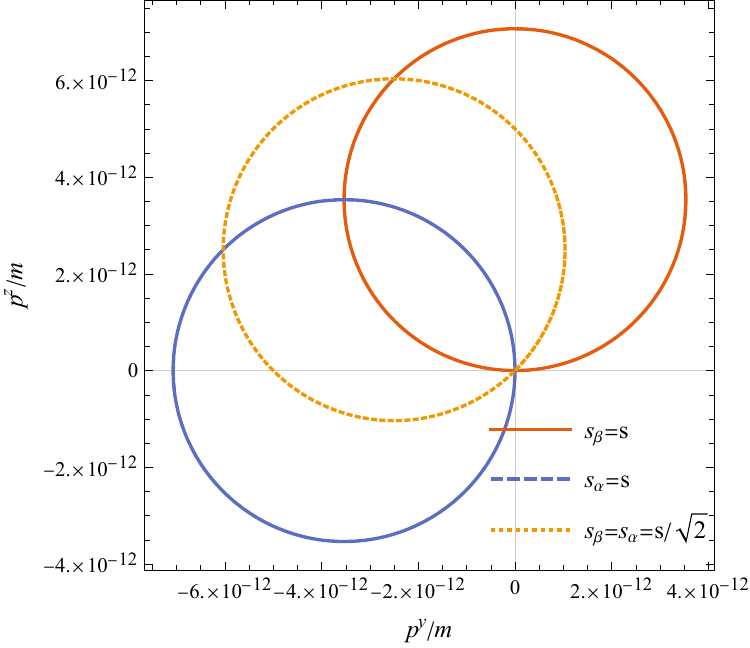}
    \caption{
    Evolution of the transverse momentum components \(p^y(u)\) and
    \(p^z(u)\) for a particle initially at rest with transverse spin
    magnitude \(s_\perp=0.1\,m\). The amplitude of the gravitational wave is \(h=10^{-10}\).
    Different colors correspond to different initial orientations in the
    \((e_2,e_3)\) plane.  
    The oscillatory structure reflects the spin--curvature coupling,
    whereas a geodesic particle would maintain \(p^y=p^z=0\).}
    \label{fig:ywithspinsy}
\end{figure}

Figure~\ref{fig:ywithspinsy} displays the corresponding evolution of the
momentum components \(p^y(u)\) and \(p^z(u)\).
While the geodesic worldline maintains vanishing transverse momentum, a
nonzero transverse spin generates an oscillatory pattern whose phase and
amplitude depend on the initial spin orientation 
(e.g., \(s_\alpha=s\,,s_\beta=0\), \(s_\beta=s\,,s_\alpha=0\), or \(s_\alpha=s_\beta=s/\sqrt{2}\)).
These oscillations directly integrate to nontrivial transverse motion
via Eqs.~\eqref{eq:y-solution}--\eqref{eq:z-solution}.
Thus the transverse spin acts as a linear-order source for deviation
from geodesic motion.

\section{Discussion}
\label{sec:discussion}
In this work, we have obtained a complete analytic solution for the motion of spinning test particles at linear order in spin within plane gravitational wave backgrounds. By combining a parallel-transported tetrad with the translational Killing symmetries inherent to the geometry, our framework provides an exact, closed-form decoupling of the nonlinear MPD equations for a specific class of exact plane wave spacetimes in Rosen coordinates. Furthermore, because this exact geometric formulation seamlessly reduces to the transverse-traceless (TT) gauge in the weak-field regime ($|h| \ll 1$), our analytic integrals can be directly utilized to compute the leading-order $\mathcal{O}(h s)$ spin-curvature response of a test body to arbitrary theoretical or numerical waveform templates. This yields a highly versatile and mathematically rigorous tool for investigating spin-induced trajectory deviations.

In our previous work~\cite{Wang:2023eqj,Chen:2025tok} we investigated the
deviation of precession angles between two freely falling gyroscopes in
plane wave spacetimes and established its connection to the velocity memory
effect.  
The present analysis lays the groundwork for extending that study to include
spin--curvature coupling.  
The $\mathcal{O}(s)$ solution obtained here directly provides the worldline and
spin corrections needed to compute the first-order spin-induced modification to
the gyroscopic deviation angle.  
We expect that this coupling will generate additional, potentially observable
contributions to gyroscopic memory, especially when the initial spin has
transverse components.

General exact plane wave spacetimes possess a five-dimensional
isometry group~\cite{Blau:2002js}.  
Under the Tulczyjew--Dixon SSC these five Killing symmetries should, in
principle, yield five independent integrals of motion, potentially sufficient to
determine the full momentum--spin dynamics without restricting to
$\mathcal{O}(s)$. However, constructing closed expressions for these conserved quantities remains challenging.  
Unlike the Kerr geometry, where the existence of a Killing--Yano tensor leads
to a hidden symmetry that decouples the MPD system \cite{witzany_analytic_2024}---plane wave isometries form
a non-Abelian algebra that does not appear to admit an analogous geometric
structure. It is therefore unclear whether a suitable displacement vector or generalized
momentum map exists that would allow a similar exact decoupling beyond the
linear-spin approximation.

Beyond exact plane waves, it would be natural to investigate spin--curvature
coupling in asymptotically flat radiative spacetimes \cite{Bondi:1962px,Sachs:1962zza}.  
In that setting, the spin memory effect and gyroscopic memory correspond to
asymptotic charges of the extended BMS algebra, and the leading precession
induced by gravitational radiation appears at order $\mathcal{O}(h^{2})$
\cite{Seraj:2022qyt}.
Whether the spin--curvature coupling contributes to observable effects at order
$\mathcal{O}(hs)$ or $\mathcal{O}(h^{2}s)$ in this broader framework is an 
interesting question.  
Such contributions would represent a new type of spin-sensitive gravitational
memory effect that has not yet been systematically explored.

Future space-based detectors such as LISA, TianQin, or Taiji \cite{LISA:2017pwj,LISA:2024hlh,TianQin:2015yph,Ruan:2018tsw}
will monitor test masses in heliocentric or geocentric orbits with unprecedented
precision.  
Although the proof masses are designed to be nearly torque-free, they carry
residual angular momentum, and the spacecraft themselves possess controlled
attitude dynamics.  
It is therefore natural to ask whether the spin--curvature coupling derived here
could produce measurable perturbations in the relative trajectories of the
test masses or spacecraft. 
Even if the intrinsic spin of the proof masses is very small, the long
integration times and large baselines may enhance tiny spin-induced corrections
to detectable levels.  
The analytic MPD solutions developed in this work offer a tractable framework
for modeling such effects in realistic waveform backgrounds, and may provide a
useful ingredient for both precision instrument modeling and mock data
generation.


%
\acknowledgments
I thank Chao-Jun Feng for helpful discussions and encouragement, and also gratefully acknowledge the support of the \emph{Institute of Gravitational Wave Astronomy, Henan Academy of Sciences}, over the past year.

\bibliographystyle{plainnat}
\bibliography{planewaveMPD}

\begin{thebibliography}{45}
\providecommand{\natexlab}[1]{#1}
\providecommand{\url}[1]{\texttt{#1}}
\expandafter\ifx\csname urlstyle\endcsname\relax
  \providecommand{\doi}[1]{doi: #1}\else
  \providecommand{\doi}{doi: \begingroup \urlstyle{rm}\Url}\fi

\bibitem[Amaro-Seoane et~al.(2017)]{LISA:2017pwj}
Pau Amaro-Seoane et~al.
\newblock {Laser Interferometer Space Antenna}.
\newblock 2 2017.

\bibitem[Andrzejewski(2025)]{andrzejewski_revisiting_2025}
K.~Andrzejewski.
\newblock Revisiting the dynamics of a charged spinning body in curved spacetime.
\newblock \emph{Classical and Quantum Gravity}, 42\penalty0 (5):\penalty0 055019, March 2025.
\newblock ISSN 0264-9381, 1361-6382.
\newblock \doi{10.1088/1361-6382/adb2d4}.
\newblock URL \url{http://arxiv.org/abs/2405.13784}.
\newblock arXiv:2405.13784 [gr-qc].

\bibitem[Bini and Damour(2017)]{Bini:2017xzy}
Donato Bini and Thibault Damour.
\newblock {Gravitational spin-orbit coupling in binary systems, post-Minkowskian approximation and effective one-body theory}.
\newblock \emph{Phys. Rev. D}, 96\penalty0 (10):\penalty0 104038, 2017.
\newblock \doi{10.1103/PhysRevD.96.104038}.

\bibitem[Bini and Felice(2000)]{bini_gyroscopes_2000}
Donato Bini and Fernando~De Felice.
\newblock Gyroscopes and gravitational waves.
\newblock \emph{Classical and Quantum Gravity}, 17\penalty0 (22):\penalty0 4627--4635, November 2000.
\newblock ISSN 0264-9381, 1361-6382.
\newblock \doi{10.1088/0264-9381/17/22/303}.
\newblock URL \url{https://iopscience.iop.org/article/10.1088/0264-9381/17/22/303}.

\bibitem[Bini et~al.(2009)Bini, Cherubini, Geralico, and Ortolan]{bini_dixons_2009}
Donato Bini, Christian Cherubini, Andrea Geralico, and Antonello Ortolan.
\newblock Dixon’s extended bodies and weak gravitational waves.
\newblock \emph{General Relativity and Gravitation}, 41\penalty0 (1):\penalty0 105--116, January 2009.
\newblock ISSN 0001-7701, 1572-9532.
\newblock \doi{10.1007/s10714-008-0657-x}.
\newblock URL \url{http://link.springer.com/10.1007/s10714-008-0657-x}.

\bibitem[Bini et~al.(2017)Bini, Geralico, and Ortolan]{bini_deviation_2017}
Donato Bini, Andrea Geralico, and Antonello Ortolan.
\newblock Deviation and precession effects in the field of a weak gravitational wave.
\newblock \emph{Physical Review D}, 95\penalty0 (10):\penalty0 104044, May 2017.
\newblock ISSN 2470-0010, 2470-0029.
\newblock \doi{10.1103/PhysRevD.95.104044}.
\newblock URL \url{http://arxiv.org/abs/1705.02794}.
\newblock arXiv:1705.02794 [gr-qc].

\bibitem[Blau and O'Loughlin(2003)]{Blau:2002js}
Matthias Blau and Martin O'Loughlin.
\newblock {Homogeneous plane waves}.
\newblock \emph{Nucl. Phys. B}, 654:\penalty0 135--176, 2003.
\newblock \doi{10.1016/S0550-3213(03)00055-5}.

\bibitem[Bondi et~al.(1962)Bondi, van~der Burg, and Metzner]{Bondi:1962px}
H.~Bondi, M.~G.~J. van~der Burg, and A.~W.~K. Metzner.
\newblock {Gravitational waves in general relativity. 7. Waves from axisymmetric isolated systems}.
\newblock \emph{Proc. Roy. Soc. Lond. A}, 269:\penalty0 21--52, 1962.
\newblock \doi{10.1098/rspa.1962.0161}.

\bibitem[Chen et~al.(2025{\natexlab{a}})Chen, Hsieh, and Lee]{Chen:2025ncm}
Yi-Ping Chen, Tien Hsieh, and Da-Shin Lee.
\newblock {Motion of spinning particles in the Kerr-Newman black hole exterior. I. Periodic orbits}.
\newblock 10 2025{\natexlab{a}}.

\bibitem[Chen et~al.(2025{\natexlab{b}})Chen, Wang, and Feng]{Chen:2025tok}
Yingxin Chen, Ke~Wang, and Chao-Jun Feng.
\newblock {Higher order analysis of the gravitational wave velocity memory effect between two free-falling gyroscopes in the plane wave spacetime}.
\newblock \emph{Phys. Rev. D}, 111\penalty0 (10):\penalty0 104085, 2025{\natexlab{b}}.
\newblock \doi{10.1103/38k7-pwlr}.

\bibitem[Ciou et~al.(2025)Ciou, Hsieh, and Lee]{Ciou:2025ygb}
Siang-Yao Ciou, Tien Hsieh, and Da-Shin Lee.
\newblock {Dynamics of spinning particles in Reissner-Nordstr{\"o}m black hole exterior}.
\newblock \emph{JCAP}, 05:\penalty0 086, 2025.
\newblock \doi{10.1088/1475-7516/2025/05/086}.

\bibitem[Colpi et~al.(2024)]{LISA:2024hlh}
Monica Colpi et~al.
\newblock {LISA Definition Study Report}.
\newblock 2 2024.

\bibitem[Costa et~al.(2018)Costa, Lukes-Gerakopoulos, and Semer{\'a}k]{Costa:2017kdr}
L.~Filipe~O. Costa, Georgios Lukes-Gerakopoulos, and Old{\v{r}}ich Semer{\'a}k.
\newblock {Spinning particles in general relativity: Momentum-velocity relation for the Mathisson-Pirani spin condition}.
\newblock \emph{Phys. Rev. D}, 97\penalty0 (8):\penalty0 084023, 2018.
\newblock \doi{10.1103/PhysRevD.97.084023}.

\bibitem[Dietrich et~al.(2021)Dietrich, Hinderer, and Samajdar]{Dietrich:2020eud}
Tim Dietrich, Tanja Hinderer, and Anuradha Samajdar.
\newblock {Interpreting Binary Neutron Star Mergers: Describing the Binary Neutron Star Dynamics, Modelling Gravitational Waveforms, and Analyzing Detections}.
\newblock \emph{Gen. Rel. Grav.}, 53\penalty0 (3):\penalty0 27, 2021.
\newblock \doi{10.1007/s10714-020-02751-6}.

\bibitem[Dixon(1970)]{dixon_dynamics_1970}
W.~G. Dixon.
\newblock Dynamics of extended bodies in general relativity. {I}. {Momentum} and angular momentum.
\newblock \emph{Proceedings of the Royal Society of London. A. Mathematical and Physical Sciences}, 314\penalty0 (1519):\penalty0 499--527, January 1970.
\newblock ISSN 0080-4630.
\newblock \doi{10.1098/rspa.1970.0020}.
\newblock URL \url{https://royalsocietypublishing.org/doi/10.1098/rspa.1970.0020}.

\bibitem[Einstein and Rosen(1937)]{Einstein:1937qu}
Albert Einstein and N.~Rosen.
\newblock {On Gravitational waves}.
\newblock \emph{J. Franklin Inst.}, 223:\penalty0 43--54, 1937.
\newblock \doi{10.1016/S0016-0032(37)90583-0}.

\bibitem[Everitt et~al.(2011)Everitt, DeBra, Parkinson, Turneaure, Conklin, Heifetz, Keiser, Silbergleit, Holmes, Kolodziejczak, Al-Meshari, Mester, Muhlfelder, Solomonik, Stahl, Worden, Bencze, Buchman, Clarke, Al-Jadaan, Al-Jibreen, Li, Lipa, Lockhart, Al-Suwaidan, Taber, and Wang]{everitt_gravity_2011}
C.~W.~F. Everitt, D.~B. DeBra, B.~W. Parkinson, J.~P. Turneaure, J.~W. Conklin, M.~I. Heifetz, G.~M. Keiser, A.~S. Silbergleit, T.~Holmes, J.~Kolodziejczak, M.~Al-Meshari, J.~C. Mester, B.~Muhlfelder, V.~G. Solomonik, K.~Stahl, P.~W. Worden, W.~Bencze, S.~Buchman, B.~Clarke, A.~Al-Jadaan, H.~Al-Jibreen, J.~Li, J.~A. Lipa, J.~M. Lockhart, B.~Al-Suwaidan, M.~Taber, and S.~Wang.
\newblock Gravity {Probe} {B}: {Final} {Results} of a {Space} {Experiment} to {Test} {General} {Relativity}.
\newblock \emph{Physical Review Letters}, 106\penalty0 (22):\penalty0 221101, May 2011.
\newblock ISSN 0031-9007, 1079-7114.
\newblock \doi{10.1103/PhysRevLett.106.221101}.
\newblock URL \url{https://link.aps.org/doi/10.1103/PhysRevLett.106.221101}.

\bibitem[Flanagan et~al.(2019)Flanagan, Grant, Harte, and Nichols]{Flanagan:2018yzh}
{\'E}anna~{\'E}. Flanagan, Alexander~M. Grant, Abraham~I. Harte, and David~A. Nichols.
\newblock {Persistent gravitational wave observables: general framework}.
\newblock \emph{Phys. Rev. D}, 99\penalty0 (8):\penalty0 084044, 2019.
\newblock \doi{10.1103/PhysRevD.99.084044}.

\bibitem[Flanagan et~al.(2020)Flanagan, Grant, Harte, and Nichols]{Flanagan:2019ezo}
{\'E}anna~{\'E}. Flanagan, Alexander~M. Grant, Abraham~I. Harte, and David~A. Nichols.
\newblock {Persistent gravitational wave observables: Nonlinear plane wave spacetimes}.
\newblock \emph{Phys. Rev. D}, 101\penalty0 (10):\penalty0 104033, 2020.
\newblock \doi{10.1103/PhysRevD.101.104033}.

\bibitem[Garriga and Verdaguer(1991)]{Garriga:1990dp}
Jaume Garriga and Enric Verdaguer.
\newblock {Scattering of quantum particles by gravitational plane waves}.
\newblock \emph{Phys. Rev. D}, 43:\penalty0 391--401, 1991.
\newblock \doi{10.1103/PhysRevD.43.391}.

\bibitem[Harte(2020)]{harte_extended-body_2020}
Abraham~I. Harte.
\newblock Extended-body motion in black hole spacetimes: {What} is possible?
\newblock \emph{Physical Review D}, 102\penalty0 (12):\penalty0 124075, December 2020.
\newblock ISSN 2470-0010, 2470-0029.
\newblock \doi{10.1103/PhysRevD.102.124075}.
\newblock URL \url{http://arxiv.org/abs/2011.00110}.
\newblock arXiv:2011.00110 [gr-qc].

\bibitem[Harte et~al.(2025)Harte, Mieling, Oancea, and Steininger]{Harte:2024mwj}
Abraham~I. Harte, Thomas~B. Mieling, Marius~A. Oancea, and Elisabeth Steininger.
\newblock {Gravitational wave memory and its effects on particles and fields}.
\newblock \emph{Phys. Rev. D}, 111\penalty0 (2):\penalty0 024034, 2025.
\newblock \doi{10.1103/PhysRevD.111.024034}.

\bibitem[Kessari et~al.(2002)Kessari, Singh, Tucker, and Wang]{Kessari:2002jc}
S.~Kessari, D.~Singh, R.~W. Tucker, and C.~Wang.
\newblock {Scattering of spinning test particles by plane gravitational and electromagnetic waves}.
\newblock \emph{Class. Quant. Grav.}, 19:\penalty0 4943--4952, 2002.
\newblock \doi{10.1088/0264-9381/19/19/312}.

\bibitem[Luo et~al.(2016)]{TianQin:2015yph}
Jun Luo et~al.
\newblock {TianQin: a space-borne gravitational wave detector}.
\newblock \emph{Class. Quant. Grav.}, 33\penalty0 (3):\penalty0 035010, 2016.
\newblock \doi{10.1088/0264-9381/33/3/035010}.

\bibitem[Marck(1983)]{marck_parallel-tetrad_1983}
Jean-Alain Marck.
\newblock Parallel-tetrad on null geodesics in {Kerr}-{Newman} space-time.
\newblock \emph{Physics Letters A}, 97\penalty0 (4):\penalty0 140--142, August 1983.
\newblock ISSN 03759601.
\newblock \doi{10.1016/0375-9601(83)90197-4}.
\newblock URL \url{https://linkinghub.elsevier.com/retrieve/pii/0375960183901974}.

\bibitem[Mohseni and Sepangi(2000)]{mohseni_gravitational_2000}
M.~Mohseni and H.~R. Sepangi.
\newblock Gravitational waves and spinning test particles.
\newblock \emph{Classical and Quantum Gravity}, 17\penalty0 (22):\penalty0 4615--4625, November 2000.
\newblock ISSN 0264-9381, 1361-6382.
\newblock \doi{10.1088/0264-9381/17/22/302}.
\newblock URL \url{http://arxiv.org/abs/gr-qc/0009070}.
\newblock arXiv:gr-qc/0009070.

\bibitem[Mohseni et~al.(2001)Mohseni, Tucker, and Wang]{mohseni_motion_2001}
M.~Mohseni, Robin~W. Tucker, and Charles Wang.
\newblock On the motion of spinning test particles in plane gravitational waves.
\newblock \emph{Classical and Quantum Gravity}, 18\penalty0 (15):\penalty0 3007--3017, August 2001.
\newblock ISSN 0264-9381, 1361-6382.
\newblock \doi{10.1088/0264-9381/18/15/314}.
\newblock URL \url{http://arxiv.org/abs/gr-qc/0308042}.
\newblock arXiv:gr-qc/0308042.

\bibitem[Mohseni(2002)]{Mohseni:2002zg}
Morteza Mohseni.
\newblock {Spinning particles in gravitational wave space-time}.
\newblock \emph{Phys. Lett. A}, 301:\penalty0 382--388, 2002.
\newblock \doi{10.1016/S0375-9601(02)00988-X}.

\bibitem[Papapetrou(1951)]{Papapetrou:1951pa}
Achille Papapetrou.
\newblock {Spinning test particles in general relativity. 1.}
\newblock \emph{Proc. Roy. Soc. Lond. A}, 209:\penalty0 248--258, 1951.
\newblock \doi{10.1098/rspa.1951.0200}.

\bibitem[Penrose(1976)]{Penrose:1976}
Roger Penrose.
\newblock {Any Space-Time has a Plane Wave as a Limit}.
\newblock \emph{Differential Geometry and Relativity}, pages 271--275, 1976.
\newblock \doi{10.1007/978-94-010-1508-0-23}.

\bibitem[Poisson et~al.(2011)Poisson, Pound, and Vega]{poisson_motion_2011}
Eric Poisson, Adam Pound, and Ian Vega.
\newblock The motion of point particles in curved spacetime.
\newblock \emph{Living Reviews in Relativity}, 14\penalty0 (1):\penalty0 7, December 2011.
\newblock ISSN 2367-3613, 1433-8351.
\newblock \doi{10.12942/lrr-2011-7}.
\newblock URL \url{http://arxiv.org/abs/1102.0529}.
\newblock arXiv:1102.0529 [gr-qc].

\bibitem[Ruan et~al.(2020)Ruan, Guo, Cai, and Zhang]{Ruan:2018tsw}
Wen-Hong Ruan, Zong-Kuan Guo, Rong-Gen Cai, and Yuan-Zhong Zhang.
\newblock {Taiji program: Gravitational-wave sources}.
\newblock \emph{Int. J. Mod. Phys. A}, 35\penalty0 (17):\penalty0 2050075, 2020.
\newblock \doi{10.1142/S0217751X2050075X}.

\bibitem[Rudiger(1981)]{rudiger_conserved_1981}
R.~Rudiger.
\newblock Conserved quantities of spinning test particles in general relativity. {I}.
\newblock \emph{Proceedings of the Royal Society of London. A. Mathematical and Physical Sciences}, 375\penalty0 (1761):\penalty0 185--193, March 1981.
\newblock ISSN 0080-4630.
\newblock \doi{10.1098/rspa.1981.0046}.
\newblock URL \url{https://royalsocietypublishing.org/doi/10.1098/rspa.1981.0046}.

\bibitem[Rudiger(1983)]{rudiger_conserved_1983}
R.~Rudiger.
\newblock Conserved quantities of spinning test particles in general relativity. {II}.
\newblock \emph{Proceedings of the Royal Society of London. A. Mathematical and Physical Sciences}, 385\penalty0 (1788):\penalty0 229--239, January 1983.
\newblock ISSN 0080-4630.
\newblock \doi{10.1098/rspa.1983.0012}.
\newblock URL \url{https://royalsocietypublishing.org/doi/10.1098/rspa.1983.0012}.

\bibitem[Sachs(1962)]{Sachs:1962zza}
R.~Sachs.
\newblock {Asymptotic symmetries in gravitational theory}.
\newblock \emph{Phys. Rev.}, 128:\penalty0 2851--2864, 1962.
\newblock \doi{10.1103/PhysRev.128.2851}.

\bibitem[Semerak(1999)]{semerak_spinning_1999}
O.~Semerak.
\newblock Spinning test particles in a {Kerr} field -- {I}.
\newblock \emph{Monthly Notices of the Royal Astronomical Society}, 308\penalty0 (3):\penalty0 863--875, September 1999.
\newblock ISSN 0035-8711, 1365-2966.
\newblock \doi{10.1046/j.1365-8711.1999.02754.x}.
\newblock URL \url{https://academic.oup.com/mnras/article/308/3/863/973826}.

\bibitem[Seraj and Oblak(2022)]{Seraj:2022qyt}
Ali Seraj and Blagoje Oblak.
\newblock {Precession Caused by Gravitational Waves}.
\newblock \emph{Phys. Rev. Lett.}, 129\penalty0 (6):\penalty0 061101, 2022.
\newblock \doi{10.1103/PhysRevLett.129.061101}.

\bibitem[Skoup{\'y} and Lukes-Gerakopoulos(2021)]{Skoupy:2021asz}
Viktor Skoup{\'y} and Georgios Lukes-Gerakopoulos.
\newblock {Spinning test body orbiting around a Kerr black hole: Eccentric equatorial orbits and their asymptotic gravitational-wave fluxes}.
\newblock \emph{Phys. Rev. D}, 103\penalty0 (10):\penalty0 104045, 2021.
\newblock \doi{10.1103/PhysRevD.103.104045}.

\bibitem[Skoupý and Witzany(2025)]{skoupy_analytic_2025}
Viktor Skoupý and Vojtěch Witzany.
\newblock Analytic {Solution} for the {Motion} of {Spinning} {Particles} in {Kerr} {Spacetime}.
\newblock \emph{Physical Review Letters}, 134\penalty0 (17):\penalty0 171401, April 2025.
\newblock ISSN 0031-9007, 1079-7114.
\newblock \doi{10.1103/PhysRevLett.134.171401}.
\newblock URL \url{https://link.aps.org/doi/10.1103/PhysRevLett.134.171401}.

\bibitem[Suzuki and Maeda(1997)]{Suzuki:1996gm}
Shingo Suzuki and Kei-ichi Maeda.
\newblock {Chaos in Schwarzschild space-time: The motion of a spinning particle}.
\newblock \emph{Phys. Rev. D}, 55:\penalty0 4848--4859, 1997.
\newblock \doi{10.1103/PhysRevD.55.4848}.

\bibitem[Tod et~al.(1976)Tod, de~Felice, and Calvani]{Tod:1976ud}
K.~P. Tod, F.~de~Felice, and M.~Calvani.
\newblock {Spinning test particles in the field of a black hole}.
\newblock \emph{Nuovo Cim. B}, 34:\penalty0 365, 1976.
\newblock \doi{10.1007/BF02728614}.

\bibitem[Tulczyjew(1959)]{Tulczyjew:1959}
W.~Tulczyjew.
\newblock {Motion of multipole particles in general relativity theory}.
\newblock \emph{Acta Phys.Polon.}, 18:\penalty0 393, 1959.

\bibitem[Wang and Feng(2023)]{Wang:2023eqj}
Ke~Wang and Chao-Jun Feng.
\newblock {Spin vector deviation and the gravitational wave memory effect between two free-falling gyroscopes in the plane wave spacetimes}.
\newblock \emph{Phys. Rev. D}, 107\penalty0 (8):\penalty0 084044, 2023.
\newblock \doi{10.1103/PhysRevD.107.084044}.

\bibitem[Wang and Feng(2024)]{Wang:2024dmn}
Ke~Wang and Chao-Jun Feng.
\newblock {Geometric deformation and redshift structure caused by plane gravitational waves}.
\newblock \emph{Phys. Lett. B}, 855:\penalty0 138875, 2024.
\newblock \doi{10.1016/j.physletb.2024.138875}.

\bibitem[Witzany and Piovano(2024)]{witzany_analytic_2024}
Vojtěch Witzany and Gabriel~Andres Piovano.
\newblock Analytic {Solutions} for the {Motion} of {Spinning} {Particles} near {Spherically} {Symmetric} {Black} {Holes} and {Exotic} {Compact} {Objects}.
\newblock \emph{Physical Review Letters}, 132\penalty0 (17):\penalty0 171401, April 2024.
\newblock ISSN 0031-9007, 1079-7114.
\newblock \doi{10.1103/PhysRevLett.132.171401}.
\newblock URL \url{https://link.aps.org/doi/10.1103/PhysRevLett.132.171401}.

\end{thebibliography}

\end{document}